\documentclass{autart}

\journal{Physics Letters A}

\usepackage{amssymb,amsmath}
\usepackage{graphicx}
\newcommand\beq{\begin{equation}}
\newcommand\eeq{\end{equation}}
\newcommand\beqa{\begin{eqnarray}}
\newcommand\eeqa{\end{eqnarray}}

\newcommand{\fk}{\widetilde{f}}

\newcommand{\rhog}{\widehat{\rho}}
\newcommand{\etag}{\widehat{\eta}}
\newcommand{\ff}{\text{f}}
\newcommand{\mm}{\text{m}}
\newcommand{\ms}{\text{ms}}

\begin{document}
\begin{frontmatter}


\title{
The penetrable-sphere fluid  in the high-temperature, high-density limit
}

\author{Luis Acedo$^\dagger$} and
\ead{luiacrod@imm.upv.es}
\author{Andr\'es Santos\corauthref{cor1}}
\ead{andres@unex.es}
\ead[url]{http://www.unex.es/fisteor/andres/}
\corauth[cor1]{Corresponding author.
\\
$^\dagger$ Present address: Instituto Universitario de Matem\'atica Multidisciplinar, Universitat Polit\`ecnica de Val\`encia,
46022 Valencia, Spain}
\address{
Departamento de F\'{\i}sica, Universidad de Extremadura,
06071 Badajoz, Spain
}

\date{\today}

\begin{abstract}
We consider a fluid of $d$-dimensional spherical particles interacting via a
pair potential $\phi(r)$ which takes a finite value $\epsilon$ if the two spheres
are overlapped ($r<\sigma$) and $0$ otherwise. This penetrable-sphere model  has been proposed to describe the effective interaction of micelles in a solvent. We derive the structural and thermodynamic functions in the limit where  the
reduced temperature $k_BT/\epsilon$ and density $\rho\sigma^d$ tend
to infinity, their ratio   being kept finite. The fluid exhibits a spinodal instability
at a certain
maximum scaled density where  the
correlation length diverges and a crystalline phase appears, even in the one-dimensional model.
By using a simple free-volume theory
for the solid phase of the model, the fluid-solid phase transition is located.
\end{abstract}

\begin{keyword}
Penetrable-sphere model \sep Soft interactions \sep Spinodal instability \sep Fluid-solid phase transition
\PACS{61.20.-p \sep 64.70.Dv \sep  64.60.-i \sep  61.25.Hq}
\end{keyword}
\end{frontmatter}

\maketitle

Most of the theoretical studies and numerical applications of the theory
of  liquids in equilibrium is devoted to particles which interact according to
unbounded spherically symmetric pair potentials
\cite{HM86,M92}. Atomic and molecular fluids have been usually
modeled in this way and a vast effort was done during the second
half of the past century in order to understand systems such as hard
spheres, the square-well model, or the Lennard--Jones liquid. The
development of the integral equations of the theory of liquids,
the Percus--Yevick (PY) approximation being among the most widely used
in this context, and their solution for some simple
models \cite{HM86,M92,WT63} were landmarks in the history of the
theory of liquids.

In the last decade, the properties of fluids of particles
interacting via \textit{bounded} pair potentials have been the subject of
an increasing interest, the Gaussian core model \cite{SS97,LLWL01,L01} and the
penetrable-sphere model \cite{LLWL01,L01,KGRCM94,LWL98,S99,FLL00,RSWL00,SF02,KS02}
being among the most popular ones.
The motivation for the
study of fluids based upon this new class of interactions
is two-fold. First, from a fundamental point of view, they are useful to unveil the
weaknesses of standard integral equation
theories and other approximations whose validity has only been
assessed from applications to unbounded potentials. More consistent closure
approximations  arise from these studies \cite{FLL00,RSWL00}. {}From a more practical point of view,
these models have also
been proposed in order to understand the peculiar behavior of some colloidal systems,
such as micelles in a solvent or star copolymer
suspensions. The particles in these colloids are constituted
by a small core surrounded by several attached polymeric arms. As a
consequence of their structure, two or more of these particles allow
a considerable degree of overlapping with a small energy cost \cite{L01}.
An ultrasoft logarithmically divergent potential for short distances has also been proposed
to describe the effective interaction between star polymers in good solvents \cite{LLWAJAR98}.
These are a few particular cases of  systems defining what is commonly known as ``soft matter'', which
has become an active field of research with many potential physical, chemical and engineering
applications.

The penetrable-sphere (PS)   interaction potential is defined as
\begin{equation}
\phi(r)=
\left\{
\begin{array}{ll}
\epsilon,& r<\sigma\\
0,&r>\sigma,
\end{array}
\right.
\label{phir}
\end{equation}
This model   was suggested
by Marquest and Witten \cite{MW89} in the late eighties as a simple theoretical
approach to the explanation of the experimentally observed crystallization
of copolymer mesophases, where  a simple cubic solid phase  coexists with the disordered suspension.
By arguments based on the internal energy alone, these authors claimed that, under the
assumption of single-site occupancy, the
stability of the simple cubic crystal was assured in a given domain of the
phase diagram. On the other hand, density-functional theory \cite{LWL98,S99} predicts a
freezing transition to
fcc solid phases with multiply occupied lattice sites.
The existence of clusters of overlapped particles (or ``clumps'')  in the PS crystal
and  glass was already pointed out by
Klein et al.\ \cite{KGRCM94}, who also performed Monte Carlo simulations on the system.
In the fluid phase, the standard integral equation theories (e.g., PY and HNC) are not very
reliable in describing the structure of the PS fluid, especially inside the core \cite{LWL98}.
Other more sophisticated closures \cite{FLL00}, as well as Rosenfeld's fundamental-measure theory
\cite{RSWL00}, are able to
predict
the correlations functions with high precision but, on the
other hand, the signature for a spatially ordered phase [a divergent structure
factor $S({k})$ for a finite wavenumber ${k}$] is not found with these
methods.
Recently, a mixture of colloids and non-interacting polymer coils has been studied
\cite{SF02}, where the
colloid-colloid interaction is assumed to be
that of hard spheres and the colloid-polymer interaction is described by the PS model.
The inhomogeneous structure of penetrable spheres in a spherical pore has also been investigated \cite{KS02}.

In order to shed further light on the properties of the PS system, in this Letter we
focus on the high-temperature, high-density region of the phase diagram.
As we will see later, this asymptotic region is mathematically equivalent to taking
$\epsilon\to 0$ and
$\sigma\to \infty$ in a scaled way, so that we recover rigorous results first derived
by Gates \cite{G75} and by Grewe and Klein \cite{GK77} in the more general framework of a class of
Kac potentials.
At the level of the thermodynamic properties, only  zeroth-order results were given in
Ref.\ \cite{GK77}, but here we  provide the explicit first-order corrections. Moreover, we combine
the exact free energy of the PS fluid with a free-volume estimate for the free energy of the
PS crystal to obtain the freezing and melting points.

We start by observing that the Mayer function of the PS interaction is simply
\begin{equation}
f_{\text{PS}}(r)=\exp\left[-\phi(r)/k_B T\right]-1=x f_{\text{HS}}(r),
\label{fMayer}
\end{equation}
where $x\equiv 1-\e^{-\epsilon/k_BT}$
is a parameter measuring the temperature of the system,
$f_{\text{HS}}(r)=-\Theta(\sigma-r)$ is the Mayer function of a hard-sphere (HS) system
($\Theta$ being the Heaviside
step function), $k_B$ is the Boltzmann constant and $T$ is the absolute temperature.
Obviously, the PS model includes the HS fluid
($T\to 0$  or $x\to 1$) and the point-particle fluid ($T\to \infty$  or
$x\to 0$) as special limits. The latter limit is assumed at finite number density $\rho$.
However, a non-trivial limit is obtained if $x \to 0$
and $\rho \to \infty$ keeping the product (\textit{scaled} density) $\widehat{\rho}=\rho x$ finite. Although for high temperatures
one has $x\approx \epsilon/k_BT$, it is more convenient to work with $x$ rather than with $k_BT/\epsilon$ as the control parameter.

Let us consider now the exact
virial expansion of the cavity function $y(r)\equiv g(r)\exp[\phi(r)/k_BT]$,
where $g(r)$ is the radial distribution function \cite{HM86}. In the diagrammatic representation
of the virial expansion of $y(r)$, each bond in a given diagram represents a Mayer function $f_{\text{PS}}(r)=x f_{\text{HS}}(r)$. Therefore,
\begin{equation}
y(r)=1+\sum_{n=1}^\infty \rho^n x^{n+1}\sum_{m=0}^{n(n+1)/2-1}x^m y_n^{(m)}(r),
\label{yrvirial}
\end{equation}
where the HS function $y_n^{(m)}(r)$ is represented by the sum of diagrams having  2 root points (white circles), $n$ field points (black
circles) and $n+1+m$ bonds. Setting $x=1$ (zero-temperature limit), Eq.\ (\ref{yrvirial})
becomes the virial series for hard spheres. On the other hand, in the high-temperature
limit $x\to 0$ with $\rho\to\infty$ and $\widehat{\rho}=\rho x$ finite, we get
\begin{equation}
y(r)=1+x\sum_{n=1}^\infty \widehat{\rho}^n y_n^{(0)}(r)+\mathcal{O}(x^2)=
1+x w(r)+\mathcal{O}(x^2) ,
\label{yrhtlim}
\end{equation}
where the second equality defines the function $w(r)$.
The functions $y_n^{(0)}(r)$ are represented by \textit{chain}  diagrams, i.e.,
\beq
y_n^{(0)}(|\mathbf{r}_0-\mathbf{r}_{n+1}|)=\; \stackrel{0}{\circ}\!\!\text{---}\!\!\stackrel{1}{\bullet}\!\!\text{---}\!\!\stackrel{2}{\bullet}\!\!\text{---}\cdots \text{---}\!\!\stackrel{n}{\bullet}\!\!\text{---}\!\!\!\!\!\stackrel{n+1}{\circ}.
\label{linear}
\eeq
By application of the convolution theorem,  the Fourier transform of
the function $w(r)$  is given by
\begin{equation}
\label{wk}
\widetilde{w}(k)={\widehat{\rho}
[\widetilde{f}_{\text{HS}}(k)]^2}\left[{1-\widehat{\rho} \widetilde{f}_{\text{HS}}(k)}\right]^{-1},
\end{equation}
where $\widetilde{f}_{\text{HS}}(k)=-(2\pi\sigma)^{d/2} k^{-d/2} J_{d/2}(k\sigma)$ is the
Fourier transform of the HS Mayer function, $J_\nu(z)$ being the Bessel function.
Similarly, the
structure factor $S(k)$ and the direct correlation function $c(r)$ are
\beq
\label{Sk}
S(k)=\left[{1-\widehat{\rho} \widetilde{f}_{\text{HS}}(k)}\right]^{-1}+\mathcal{O}(x) ,
\eeq
\beq
{c}(r)=x {f}_{\text{HS}}(r)+\mathcal{O}(x^2) .
\label{cr}
\eeq
{}From Eqs.\ (\ref{yrhtlim}) and (\ref{cr}) it follows that 
 the PY and HNC closures become exact to first order in $x$. The same happens with  any sensible approximation
which retains the chain diagrams of the virial expansion.
It is worth noting that the non-linear Debye-H\"uckel for the radial distribution function of a system of charged particles can be derived by neglecting all but the chain diagrams, which are most weakly connected (and hence the most strongly divergent) ones \cite{HM86}. More in general, the chain diagrams determine the asymptotic long-distance behavior of the correlation functions \cite{M92}.

\begin{table}[t]
\caption{\label{table1}
Contributions of zeroth- and first-order in $x$ to the main thermodynamic quantities.}
\begin{tabular}{ccc}
\hline
Quantity&$\mathcal{O}(x^0)$&$\mathcal{O}(x)$\\
\hline
$a_{\text{ex}}/k_BT$&$2^{d-1}\widehat{\eta}$&$2^{d-1} \gamma(\sigma)$\\
$p/\rho k_B T$&$1+2^{d-1}\widehat{\eta}$&$2^{d-1}\widehat{\eta} w(\sigma)$\\
$s_{\text{ex}}/k_B$&0&$2^{d-1}\widehat{\eta}\left[w(\sigma)-\frac{1}{2}\right]$\\
$\mu_{\text{ex}}/k_BT$&$2^d\widehat{\eta}$&$2^{d-1}\left[\widehat{\eta}-2^{-d}w(0)\right]$\\
$u_{\text{ex}}/k_BT$&$2^{d-1}\widehat{\eta}$&$2^{d-2}\left[\widehat{\eta}-2^{1-d}w(0)\right]$\\
$c_{\text{ex}}/k_B$&$0$&$\frac{1}{2}\widehat{\eta}\partial w(0)/\partial\widehat{\eta}$\\
$(\partial p/\partial\rho)_T/k_BT$&$1+2^d\widehat{\eta}$&$2^{d-1}\partial[\widehat{\eta}^2w(\sigma)]/\partial\widehat{\eta}$\\
\hline
\end{tabular}
\end{table}
Besides the standard
structure functions of the theory of liquids listed above we have found it useful
in this case to introduce an auxiliary function $\gamma(r)$ as the density integral of $w(r)$, namely $w(r)=\partial \gamma(r)/\partial \etag$, where $\widehat{\eta}=\widehat{\rho} v_d \sigma^d$ is the \textit{scaled} packing fraction,
$v_d=(\pi /4)^{d/2}/\Gamma (1+d/2)$ being the volume of a $d$-dimensional
sphere of unit diameter.
In Fourier space,
\beq
\widetilde{\gamma}(k)=-\etag \fk_{\text{HS}}(k)-v_d\sigma^d \ln\left[1-\rhog \fk_{\text{HS}}(k)\right].
\label{gammak}
\eeq
We have verified that the values of $\gamma(r)$ at $r=0$ and $r=\sigma$ satisfy
the linear relation
\beq
\gamma(\sigma)+\frac{2^{-d}}{\widehat{\eta}}\gamma(0)=
\frac{\widehat{\eta}}{2},
\label{gamma0}
\eeq
which is crucial to prove the thermodynamic consistency between the virial and energy routes to the equation of state.
By standard application of statistical-mechanical formulas  relating the correlation functions to the thermodynamic quantities \cite{HM86}, we have derived expressions for the latter  to first order in $x$ in terms of the values of $w(r)$ and $\gamma(r)$ at $r=0$ and $r=\sigma$. The  (excess) free energy per particle $a_{\text{ex}}$, the compressibility factor $p/\rho k_BT$, the (excess) entropy per particle $s_{\text{ex}}$, the (excess) chemical potential $\mu_{\text{ex}}$, the (excess) internal energy per particle $u_{\text{ex}}$,  the (excess)  specific heat $c_{\text{ex}}$ and the inverse isothermal compressibility $(\partial p/\partial\rho)_T/k_BT$ are listed  in Table \ref{table1}. The second and third columns give the zeroth- and first-order contributions, respectively, in the exact expansions of those quantities in powers of $x$ at constant $\widehat{\rho}=\rho x$. The zeroth-order terms are linear functions of $\widehat{\rho}$ (or $\widehat{\eta}$), i.e., the virial expansion truncated after the second virial coefficient becomes exact in the limit $x\to 0$. However,
the first-order terms are highly nonlinear functions of the scaled density and so all the virial coefficients contribute.

Gates \cite{G75} and Grewe and Klein \cite{GK77} considered a class of Kac potentials of the form $\phi(r)=\epsilon \phi^*(r/\sigma)$, where  $\phi^*(r^*)$ is a non-negative bounded and integrable function and $\epsilon\propto \sigma^{-d}$. They proved rigorously that in the van der Waals limit $\sigma\to \infty$, the direct correlation function adopts the mean-field expression $c(r)=-\phi(r)/k_BT$ and hence the structure factor is $S(k)=\left[1+(\rho/k_BT)\widetilde{\phi}(k)\right]^{-1}$. The PS potential belongs to the Kac class with $\phi^*(r^*)=\Theta(1-r^*)$ and so in the van der Waals limit one has $x\to 0$ with $\etag$ fixed. It is then obvious that the van der Waals limit ($\epsilon\propto \sigma^{-d}$, $\sigma\to\infty$ at fixed $T$ and $\rho$) is equivalent to the high-temperature, high-density limit (at fixed $\epsilon$ and $\sigma$) considered in this paper. In fact, Eqs.\ (\ref{Sk}) and (\ref{cr}) are recovered from the more general results of Refs.\ \cite{G75} and \cite{GK77}, although the route followed here  differs from theirs.
Equations (\ref{Sk}) and (\ref{cr}) were also found by Likos et al.\ \cite{LLWL01} as a mean-field approximation in the limit $\rho\to\infty$.  However, as is apparent from our derivation of Eqs.\ (\ref{Sk}) and (\ref{cr}), the key ingredient is the high-temperature assumption. The additional high-density assumption is only needed to depart from the trivial results corresponding to a gas of non-interacting particles. Comparison with Monte Carlo simulations shows that the mean-field structural functions behave very well even for $x\lesssim 0.2$ \cite{LLWL01}.

\begin{figure}
 \includegraphics[width=1.0 \columnwidth]{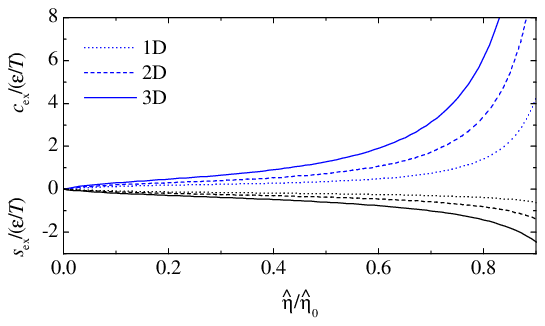}
 \caption{Plot of $s_{\text{ex}}/(\epsilon/T)$ (lower curves) and $c_{\text{ex}}/(\epsilon/T)$ (upper curves) versus $\etag/\etag_0$ for $d=1$ (dotted line), $d=2$ (dashed line) and $d=3$ (solid line).\label{entropy}}
 \end{figure}

The results for the pressure, the internal energy and the isothermal compressibility to zeroth-order in $x$ were already given by Grewe and Klein \cite{GK77}. However, Eq.\ (\ref{gamma0}) and the third column of Table \ref{table1} are, to the best of our knowledge, new results. The contributions of $\mathcal{O}(x)$ are especially relevant in the case of the excess entropy and specific heat, since the corresponding $\mathcal{O}(x^0)$-terms vanish. Figure \ref{entropy} shows the high-temperature limit of $s_{\text{ex}}/(\epsilon/T)$ and $c_{\text{ex}}/(\epsilon/T)$ versus $\etag/\etag_0$ (where $\etag_0$ will be defined below). The nonlinear density-dependence of both quantities is clearly observed.

\begin{figure}
 \includegraphics[width=1.0 \columnwidth]{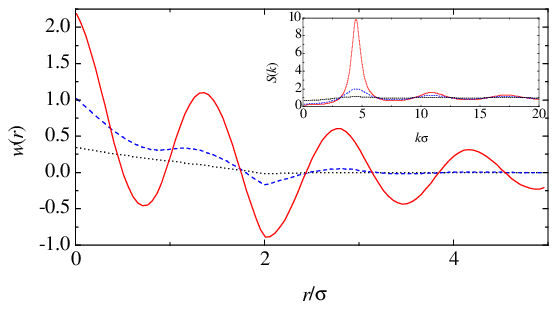}
 \caption{Plot of $w(r)$ and $S(k)$ (see inset) at $\widehat{\eta}/\widehat{\eta}_0=0.1$ (dotted line),
$0.5$ (dashed line) and $0.9$ (solid line) for $d=1$.\label{w_r}}
 \end{figure}

As analyzed  by Grewe and Klein \cite{GK77,KG80}, the PS fluid presents a spinodal point (Kirkwood instability) at a certain scaled density. A simple examination of Eqs.\ (\ref{wk}), (\ref{Sk}) and (\ref{gammak}) shows that these quantities are well defined for every real wavenumber $k$ if and only
if the maximum value  of $\widetilde{f}_{\text{HS}}(k)$ is smaller than $1/\widehat{\rho}$. In general, the absolute maximum $\widetilde{f}_{\text{max}}$ of  $\fk_{\text{HS}} (k)$ occurs at $k_{0}$, where $k_0\sigma$ is  the first
zero of $J_{d/2+1}(z)$. Therefore,  there exists an upper bound $\widehat{\rho}_{0}=1/\widetilde{f}_{\text{max}}$ to
the scaled density, such that the
structure factor becomes divergent at the wavenumber
$k=k_{0}$ when $\widehat{\rho}\to\widehat{\rho}_{0}$. The values of $k_0$ and $\etag_0=\widehat{\rho}_0 v_d \sigma^d$ for $d=1$--5 are displayed in the  third and fifth columns of Table \ref{table2}, respectively.
As an illustration, in Fig.\ \ref{w_r} we have plotted the functions $w(r)$ and $S(k)$ corresponding
to the one-dimensional PS fluid for $\widehat{\eta}=0.1\widehat{\eta}_0$, $0.5\widehat{\eta}_0$
and $0.9\widehat{\eta}_0$.
\begin{table*}
\begin{center}
\caption{\label{table2} Values of the HS close-packing fraction $\eta_{\text{cp}}$, the wavenumber $k_0$, the nearest-neighbor distance $r_0$, the (scaled) spinodal instability packing fraction $\etag_0$,  the (scaled) freezing packing fraction $\etag_\ff$, the (scaled) packing fraction $\etag_\ms$
at the condition of marginal stability and the (scaled) melting packing fraction $\etag_\mm$.}
\begin{tabular}{cccccccc}
\hline
$d$&$\eta_{\text{cp}}$&$k_0\sigma$&$r_0/\sigma$&$\widehat{\eta}_0$&$\widehat{\eta}_\ff$&$\widehat{\eta}_\ms$&$\widehat{\eta}_\mm$\\
\hline
1&1&4.49&1.40&2.30&1.00&1.00&1.00\\
2&$\sqrt{3}\pi/6\simeq 0.91$&5.14&1.37&1.89&0.90&0.95&1.01\\
3&$\sqrt{2}\pi/6\simeq 0.74$&5.76&1.34&1.45&0.63&0.69&0.77\\
4&$\pi^2/16\simeq 0.62$&6.38&1.32&1.07&0.37&0.41&0.48\\
5&$\sqrt{2}\pi^2/30\simeq 0.47$&6.99&1.30&0.76&0.22&0.26&0.32\\
\hline
\end{tabular}
\end{center}
\end{table*}

How do the structural and thermodynamic functions behave as the scaled packing fraction $\etag$ approaches its upper bound $\etag_0$ from below?  Let us denote by $k=\pm\kappa(\etag)\pm \mathrm{i}
q(\etag)$  (with the convention $\kappa,q>0$) the four zeroes of $1-\rhog\fk_{\text{HS}}(k)$ closest to the real axis in the complex $k$-plane. These values  are responsible for the
asymptotic behavior of $w(r)$ [and hence of $g(r)$] for long distances: $\kappa$ is the
wavenumber of the oscillations, while $q$ is the damping coefficient, i.e.,
$\xi=q^{-1}$ is the \textit{correlation length}. A
straightforward asymptotic analysis yields
$
\kappa\sigma\approx k_0\sigma-[(d+1)/{3k_0\sigma}](1-\etag/\etag_0)$ and
\begin{equation}
{\xi}/\sigma
=(q\sigma)^{-1}\approx {2}^{-1/2}(1-\widehat{\eta}/\widehat{\eta}_0)^{-1/2}  ,
\label{qsigma}
\end{equation}
which implies that the correlation length diverges with a critical
exponent $\nu=\frac{1}{2}$ as the density approaches the maximum density \cite{KB81}. Moreover, the correlation function  $w(r)$ behaves as $w(r) \approx (1-\widehat{\eta}/\widehat{\eta}_0)^{-1/2}
\omega(r)$, where the scaling
function $\omega(r)$ is given by
\begin{equation}
\label{omegar}
\omega(r)=\displaystyle\frac{(2\pi)^{-d/2-1}}{\sqrt{2}
\widehat{\rho}_0 \sigma^d} (k_0\sigma)^{d-1}(k_0 r)^{-d/2+1}J_{d/2-1}(k_0 r)  .
\end{equation}
So, the first-order contribution $w(r)$ to the cavity and radial distribution functions \textit{diverges}  as $\widehat{\eta}$
tends to the maximum value $\widehat{\eta}_0$.
On the other hand, the auxiliary function $\gamma(r)$ remains finite in that limit.
According to Table \ref{table1}, the first-order coefficients in $x$ of the
pressure, the entropy, the chemical potential and the  internal energy
diverge as $(1-\widehat{\eta}/\widehat{\eta}_0)^{-1/2}$ when $\widehat{\eta}\rightarrow
\widehat{\eta}_0$, while the coefficients of the specific heat and the isothermal compressibility diverge  as $(1-\widehat{\eta}/\widehat{\eta}_0)^{-3/2}$.
Equation
(\ref{omegar}) implies that $\omega(r)\sim (k_0r)^{-(d-1)/2}\cos[k_0r-(d-1)\pi/4]$ for large $k_0r$, so the correlations  at $\etag=\etag_0$ oscillate with a wavenumber $k_0$, the amplitude decaying  algebraically.
The first maximum of $\omega(r)$ (apart from the one at  $r=0$) occurs at a value $r=r_0$ such that $k_0r_0$ is the second zero of $J_{d/2}(z)$. The quantity  $r_0$ represents the distance between nearest neighbors at $\etag=\etag_0$. It is given in Table \ref{table2} for $d=1$--5.

\begin{figure}
 \includegraphics[width=1.0 \columnwidth]{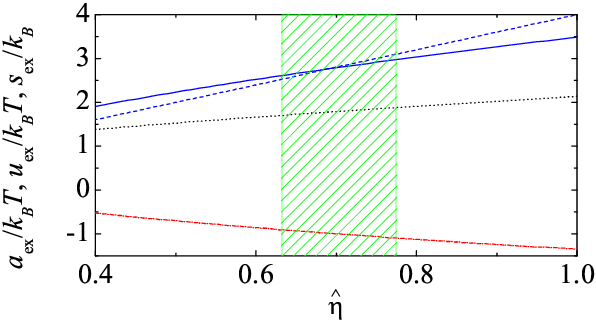}
 \caption{Excess free energy per particle in the three-dimensional
PS solid, $a^{\text{solid}}_{\text{ex}}/k_B T$ (solid line), and
 PS fluid, $a^{\text{fluid}}_{\text{ex}}/k_B T$ (dashed line),
in the high-temperature limit.
The excess internal energy, $u^{\text{solid}}_{\text{ex}}/k_B T$ (dotted line), and the excess entropy, $s^{\text{solid}}_{\text{ex}}/k_B$ (dashed-dotted line), of the PS solid are also plotted. The shaded area represents the fluid-solid coexistence region. \label{a_ex}}
 \end{figure}

On physical grounds, it is expected that  the freezing transition from the stable fluid to the stable solid   occurs at a scaled density
$\etag_\ff$ smaller than the  value $\etag_0$ at the spinodal instability. We have estimated
the values of the scaled packing fraction at freezing ($\etag_\ff$), at melting ($\etag_\mm$)
and at the point of marginal stability ($\etag_\ms$), by using for the PS solid phase a simple
free-volume theory based on the one for the HS solid \cite{LWL98,VMN99}.
In the basic picture of the PS solid \cite{L01,KGRCM94,LWL98}, the lattice sites are occupied
by ``clusters'' of
spheres that overlap each other. A particle inside one of these
clusters performs random motions but excursions beyond the
lattice distance are forbidden because of the large energy cost
associated with overlappings with  particles in the
neighbor clusters. In the high-temperature limit these
clusters are expected to contain typically a large number of particles. We will
suppose that, on average, this number scales as $\alpha/x$
(where $\alpha$ is a density-dependent parameter to be determined), so
$\eta/(\alpha/x)=\widehat{\eta}/\alpha$ is the packing fraction
of the clusters. Consequently, the excess internal energy per particle of the PS solid is
$u_{\text{ex}}^{\text{solid}}/k_BT=\alpha/2$.
Under the assumption that every cluster
behaves as a hard-core particle with a free volume $v_{\text{free}}(\etag/\alpha)$,
where
\beq
v_{\text{free}}(\eta)=
\frac{2^d}{\rho}\left[1-\left(\frac{\eta}
{\eta_{\text{cp}}}\right)^{1/d}\right]^d
\label{vfree}
\eeq
is an estimate of the free volume of a hard sphere in a crystal with packing fraction
$\eta$ \cite{LWL98,VMN99} and $\eta_{\text{cp}}$ is the HS close-packing fraction,
we have  estimated the excess entropy per particle of the PS solid as
\begin{equation}
\displaystyle\frac{s^{\text{solid}}_{\text{ex}}}{k_B}
=d\ln \left[1-\left(\frac{\widehat{\eta}}{\alpha
\eta_{\text{cp}}}\right)^{1/d}\right]+d\ln 2  .
\label{ssolid}
\end{equation}
Given a value of $\etag$, the parameter $\alpha$ is determined by minimizing the free energy
$a_{\text{ex}}^{\text{solid}}=u_{\text{ex}}^{\text{solid}}-Ts_{\text{ex}}^{\text{solid}}$.
In this way we find that $\alpha(\etag)$ is the solution of the algebraic equation
$\etag/\eta_{\text{cp}}=\alpha^{d+1}/(2+\alpha)^d$, so that
\beq
\frac{a^{\text{solid}}_{\text{ex}}(\widehat{\eta})}{k_BT}=\frac{\alpha(\etag)}{2}
+d\ln\left[\frac{1}{2}+\frac{\alpha(\etag)}{4}\right].
\label{asolid}
\eeq
The pressure and the excess chemical potential of the high-temperature PS solid are then
$p^{\text{solid}}/\rho k_BT=1+\alpha(\etag)/2$,
$\mu^{\text{solid}}_{\text{ex}}(\widehat{\eta})/{k_BT}=\alpha(\etag)+
d\ln\left[1/2+\alpha(\etag)/4\right]$.
 In
Fig.\ \ref{a_ex} we have plotted the excess free energy per
particle of the solid and the fluid phases for $d=3$. The excess internal energy and the excess entropy of the solid are also plotted.
For the fluid we have
used the zeroth-order approximation given in Table \ref{table1}, so
$a_{\text{ex}}^{\text{fluid}}(\etag)/k_BT\approx
u_{\text{ex}}^{\text{fluid}}(\etag)/k_BT\approx 2^{d-1}\etag$ and
$s_{\text{ex}}^{\text{fluid}}(\etag)/k_B\approx 0$.
While the solid has a smaller entropy than the fluid, it requires less internal energy.
As the density increases, the latter effect dominates over the former and the solid becomes
more stable than the fluid. The (scaled) density of \textit{marginal} stability $\etag_\ms$
 is determined by the
condition $a_{\text{ex}}^{\text{solid}}=a_{\text{ex}}^{\text{fluid}}$.
The (scaled) freezing and melting densities are obtained from the equality of the pressure
and the chemical potential in both phases: $p^{\text{fluid}}(\etag_\ff)=p^{\text{solid}}
(\etag_\mm)$, $\mu^{\text{fluid}}(\etag_\ff)=\mu^{\text{solid}}
(\etag_\mm)$.
 The values of $\widehat{\eta}_\mm$, $\widehat{\eta}_\ms$ and $\widehat{\eta}_\ff$
  are
given in Table \ref{table2} for $d=1$--$5$. We observe that our heuristic estimates for the
characteristic densities of the fluid-solid transition are
 typically less than half the upper bound  density $\etag_0$.

The one-dimensional (1D) case  deserves some special comments. For $d=1$, the curves representing
 the free energies of the solid and the fluid do not cross,  but  ``kiss'' each other at $\etag=1$ (i.e., they have the same value and slope at  $\etag=1$). In fact,   $a_{\text{ex}}^{\text{solid}}(\etag)<a_{\text{ex}}^{\text{fluid}}(\etag)$ not only for $\etag>1$ but also for $\etag<1$. This ``exaggerated'' stability of the 1D PS solid for small densities is obviouly an artifact of the heuristic free-volume theory we have employed. Nevertheless, our free-volume theory contains the basic ingredients explaining that the high-temperature 1D PS solid becomes more stable than the fluid at sufficiently large densities $\etag$. This is further confirmed by the rigorous existence of the spinodal instability at $\etag_0\simeq 2.30$, where the metastable fluid ceases to exist and a continuous transition to a crystalline solid takes place.
 We must emphasize that the fluid-solid phase transition in the 1D  PS model is not forbidden by van Hove's
theorem \cite{vH50} because one of the hypothesis of the
theorem, requiring the interaction potential to include an infinite
repulsive core that does not allow full overlaps between particles, is not fulfilled
\cite{CS02}. Thus, the PS model provides one of the rare examples of one-dimensional models exhibiting phase transitions \cite{CS02}. Whether the fluid-solid transition occurs near $\etag=1$ or whether it does not present a density jump, as indicated in Table \ref{table2}, are questions that need further theoretical and  simulational work before being satisfactorily elucidated.

In this paper we have focused on the high-temperature domain ($x\to 0$) of the $d$-dimensional PS model. When the exact diagrammatic expansion of the cavity function is considered,  it turns out that $x$ acts as an ordering parameter, so that the first-order term
contains only chain diagrams, which can be easily resummed, yielding mean-field expressions. If the unscaled packing fraction $\eta$ is kept finite, one arrives at the trivial case of non-interacting particles. If, on the other hand, one explores the high-density regime $\eta=\etag/x$, much more interesting results appear. To zeroth-order in $x$ \cite{GK77} the thermodynamic quantities are described by the second virial coefficient alone, but the first-order corrections exhibit a rich nonlinear dependence on $\etag$. The fluid presents a spinodal instability at the upper bound density $\eta_0(x)=\etag_0/x$,
but this is preempted by a first-order phase transition to the solid at the freezing density $\eta_\ff(x)=\etag_\ff/x$.
 It seems natural to expect that a similar situation applies for finite and low temperatures, except that the $x$-dependence of $\eta_0(x)$ and $\eta_\ff (x)$ is more complicated than in the high-temperature limit. {}From that point of view, it can be conjectured that the zero-temperature limit (where the PS model reduces to the HS model) of the upper bound density is $\lim_{x\to 1}\eta_0(x)=\eta_{\text{cp}}$. Analogously, $\lim_{x\to 1}\eta_\ff(x)=\eta_\ff^{\text{HS}}$, where $\eta_\ff^{\text{HS}}$ is the freezing packing fraction of the HS fluid.
Since $\eta_{\text{cp}}<\etag_0$  (cf.\ Table \ref{table2}) and $\eta_\ff^{\text{HS}}<\etag_\ff$, it seems plausible that the  products $x\eta_0(x)$ and $x\eta_\ff(x)$  are smoothly decreasing functions of  $x$ bounded between the values $\etag_0$ and $\etag_\ff$, respectively, at $x=0$ and the values $\eta_{\text{cp}}$ and $\eta_\ff^{\text{HS}}$, respectively, at $x=1$.
This suggests the possibility of  constructing a simple approximate theory for the PS model spanning the whole temperature spectrum, by interpolating between the successful PY theory for hard spheres (zero temperature) and the mean-field-theory results (high temperatures).
We are currently working along these lines and further results
will be published elsewhere.

We are grateful to  J.A. Cuesta, J.W. Dufty, E. Lomba, A. Malijevsk\'y, A. S\'anchez and
M. Silbert for useful discussions about the topic of this Letter. This work has been partially supported by the
Ministerio de
Ciencia y Tecnolog\'{\i}a
 (Spain) through grant No.\ BFM2001-0718 and  by the European
Community's Human Potential Programme under contract HPRN-CT-2002-00307,
DYGLAGEMEM.

\end{document}